# Geometry-Encoded Multireceiver Fluorometry Affords Full-Range Nonlinear Quantification under the Primary Inner Filter Effect

*Shuiyi Tan[2], Yuchao Fu[1,3*], Nai-Quan Zhu[3,4], Olivier J. F. Martin[1]*

[1] *Nanophotonics and Metrology Laboratory, École polytechnique fédérale de Lausanne (EPFL), Lausanne, Switzerland*

[2] *Department of Hematology, Huashan Hospital, Fudan University, Shanghai, China*

[3] *School of Electronic Information and Electrical Engineering, Shanghai Jiao Tong University, Shanghai, China*

[4] *Cinbio, Universidade de Vigo, Vigo, Spain*

## ABSTRACT

The inner filter effect (IFE) transforms the nominally linear fluorescence–concentration relationship into a geometry-dependent and often nonmonotonic response, resulting in reduced sensitivity, concentration ambiguity, and systematic underestimation at high optical densities. Here, we introduce a spatially encoded multireceiver fluorometric strategy that does not eliminate or correct the IFE, but instead harnesses the spatial fluorescence attenuation induced by IFE as an additional quantitative encoding dimension. Fluorescence generated along the excitation axis is integrated over independently positioned receiver windows, and concentration is recovered by nonlinear optimization of the joint fluorescence intensity vector. Tryptophan was selected as a biomedically relevant model fluorophore to validate the proposed strategy. Single-window calibration exhibited vanishing-gradient boundary and two-valued concentration inversions, whereas two spatially separated receiver windows restored global identifiability across the full concentration range. Screening of 35 two-receiver geometries identified the minimum-uncertainty configuration, achieving an average 95% error half-width of $\overline{E_{95}} = 0.884\ \mathrm{mg/L}$, and the maximum-sensitivity configuration, reaching a noise-normalized response sensitivity of $\overline{S_N} = 3.832\ \mathrm{a.u./(mg/L)}$. Continuous genetic optimization further converged to closely related geometric solutions, confirming the robustness of the identified design space. Across 125 experimental measurements spanning $5 - 125\ \mathrm{mg/L}$, multireceiver inversion achieved $R^2$ values of $0.997$, reduced $RMSE$ to $1.934\ \mathrm{mg/L}$ and $MAE$ to $1.440\ \mathrm{mg/L}$. By converting spatial attenuation into a multidimensional concentration coordinate, this approach extends quantitative fluorescence analysis without dilution, a separate absorbance measurement, or piecewise calibration and provides a general metrological framework for fluorescence quantification even under strong IFE conditions.

## INTRODUCTION

Fluorescence measurements are attractive for biochemical assays, clinical diagnostics, molecular imaging, and point-of-care testing because they combine high sensitivity with flexible molecular recognition and compact instrumentation [1-6]. The familiar proportionality between fluorescence intensity and fluorophore concentration, however, is a dilute-limit approximation. Attenuation of the incident beams before it reaches the observation volume produces the primary IFE (pIFE) [7-9], whereas reabsorption of emitted photons on their way to the detector produces the secondary IFE (sIFE) [9-11]. Both depend on the absorption spectrum and on the illumination and collection geometry. sIFE is especially consequential when absorption and emission overlap [11]. Consequently, a response can change from nearly linear to sublinear, reach a stationary point (vanishing-gradient point), and finally decrease as concentration increases [7, 12, 13]. A decreasing high-concentration signal is dangerous wherever a scalar optical response is interpreted as a unique concentration.

The high-dose hook effect of sandwich immunoassays creates the same metrological failure mode - a high analyte burden can generate a deceptively low signal [14, 15]. False-low results have been documented for prostate-specific antigen, hCG, and lateral-flow immunoassays, including substantial platform-dependent differences in the concentration at which the downturn appears [14, 15]. Fluorogenic thrombin-generation assays present a closer optical example, because accumulation of fluorescent product can make the measured signal nonlinear and require artifact correction [16, 17]. These cases motivate analytical designs that explicitly assess global uniqueness and accuracy, rather than assuming a monotonic quantitative response. Protein fluorescence is dominated by tryptophan near 280 nm excitation and approximately 350 nm emission, enabling label-free protein assays and conformational studies [18, 19]. Yet ultraviolet absorbance rises rapidly with protein or free-tryptophan concentration, and recent measurements of concentrated therapeutic antibodies identify pIFE as a principal obstacle to interpreting intrinsic tryptophan fluorescence [19]. In plasma and flowing blood, Indocyanine green (ICG) fluorescence increases only to a concentration-dependent maximum and then decreases, the effect complicates quantitative interpretation in angiography and fluorescence-guided procedures [20, 21]. Thus, optical nonmonotonicity is not confined to ideal dye solutions.

Most IFE methods attempt to reconstruct the fluorescence that would have been measured in an optically dilute sample. The classical midpoint correction uses absorbance at the excitation and emission wavelengths [8, 13], whereas cell-shift, mirror-cell, water-Raman, and observation-field methods estimate effective excitation and emission path lengths [22-26]. Recent approaches exploit microplate focal position, Rayleigh scattering, in-line calibration, an added absorber, or an explicit emission-angle model [12, 27-30]. These advances are important: effective-geometry correction has extended linearity to high absorbance, variable-$z$ microplate measurements have removed the need for a separate absorbance reading, and the AddAbs method has reported linear response across exceptionally concentrated samples [12, 23, 28]. Nevertheless, correction validity remains conditional on optical geometry, spectral stability, scattering, and separation of IFE from true concentration quenching or aggregation. Dilution can perturb equilibria, while auxiliary absorbance, added chromophores, or instrument-specific calibration increase workflow complexity [28, 31-33]. An alternative is to shorten the optical path or observe the sample close to the illuminated face [2]. Front-face and microscopic-domain measurements reduce attenuation and have enabled work with opaque or highly absorbing samples [10]. Spatially combined observation segments have also been shown to extend fluorescence working range [34]. These studies suggest a deeper possibility: the attenuation gradient itself is concentration dependent and therefore contains information. If several receiver windows integrate different portions of that gradient, their joint fluorescent signal vector can remain one-to-one with concentration even when every scalar channel is nonmonotonic.

Here, we implement this geometry-encoded principle using a multireceiver fluorometric configuration with lateral, orthogonal (90°) collection (Figure 1A). The position and spatial extent of each receiver can be defined independently, thereby making the observation geometry an optimizable analytical variable. We derive a spatial forward model for the inner filter effect, formulate a nonlinear inversion framework that accounts for both error tolerance and response sensitivity, quantify concentration precision using a profile-likelihood error width, and optimize the receiver geometry through both discrete configuration screening and a continuous genetic algorithm. Three nonlinear optimization strategies are evaluated for multireceiver fluorescence quantification. Experimental measurements of tryptophan fluorescence demonstrate that both two- and three-channel spatial configurations enable unique and full-range concentration quantification through the anomalous high-concentration branch.

## THEORY AND MEASUREMENT PRINCIPLE

Let $c$ be analyte concentration and $\varepsilon_{\text{ex}}$ the decadic molar absorption coefficient at the excitation wavelength. The corresponding Napierian attenuation coefficient is

$$\mu_{\text{ex}} = \ln(10)\varepsilon_{\text{ex}}c \tag{1}$$

For a collimated excitation beam propagating along $x$, conservation of radiant intensity gives

$$dI_{\text{ex}} = -\mu_{\text{ex}}I_{\text{ex}}dx \tag{2}$$

with the solution

$$I_{\text{ex}}(x) = I_0 e^{-\mu_{\text{ex}}x} \tag{3}$$

In the weak-absorption limit, $\mu_{\text{ex}}x \ll 1$, a first-order Taylor expansion gives the Beer–Lambert attenuation law [7, 35]. At finite optical density, the excitation field is intrinsically nonuniform.

For a thin slab from $x$ to $x + \Delta x$, absorbed excitation is the difference between entering and exiting intensities,

$$\Delta I_{\text{abs}}(x) = I_{\text{ex}}(x) - I_{\text{ex}}(x + \Delta x) \tag{4}$$

which, in the differential limit ($dI_{\text{ex}}(x) = I_{\text{ex}}(x + dx) - I_{\text{ex}}(x)$), becomes

$$dI_{\text{abs}}(x) = \mu_{\text{ex}}I_{\text{ex}}(x)dx \tag{5}$$

Therefore, under the spatial attenuation of the excitation light, the amount of excitation light absorbed per unit length is given by

$$\frac{dI_{\text{abs}}}{dx} = I_0\mu_{\text{ex}}e^{-\mu_{\text{ex}}x} \tag{6}$$

Let $\Phi_f$ be the fluorescence quantum yield and $K$ combine collection solid angle, optical throughput, detector responsivity, and wavelength-independent scale factors. Fluorescence generated in the slab $dx$ is

$$dI_f(x) = K\Phi_f dI_{\text{abs}}(x) \tag{7}$$

so, the fluorescence-generation line density (defined as $f(x) = \frac{dI_f}{dx}$) is

$$f(x) = K\Phi_f I_0\mu_{\text{ex}}e^{-\mu_{\text{ex}}x} \tag{8}$$

The extended sample cell in Figure 1B directly visualizes this pIFE gradient: fluorescence is strongest near the excitation entrance and decays more steeply as concentration increases.

To include sIFE, define the emission-wavelength Napierian attenuation coefficient

$$\mu_{\text{em}} = \ln(10)\varepsilon_{\text{em}}c \tag{9}$$

and let $l_{\text{em}}(x)$ be the path from a fluorescence-generating position to the detector. The emission transmittance is

$$T_{\text{em}}(x) = e^{-\mu_{\text{em}}l_{\text{em}}(x)} \tag{10}$$

giving the general detected line density

$$f_{\text{det}}(x) = K\Phi_f I_0\mu_{\text{ex}}e^{-\mu_{\text{ex}}x-\mu_{\text{em}}l_{\text{em}}(x)} \tag{11}$$

This form separates excitation-depth attenuation from the geometry-dependent emission path. Lateral, forward, and backward collection are derived explicitly in Supporting Information Section S1 [27, 36]. For the lateral orthogonal geometric configuration used in this work, $l_{\text{em}} = l_y$ is approximately independent of $x$ over an aperture-defined window. Receiver $j$ spanning $[x_{1j}, x_{2j}]$ records

$$I_{f,j}(c) = \int_{x_{1j}}^{x_{2j}} f_{\text{det}}(x)\, dx \tag{12}$$

and analytical integration yields

$$I_{f,j}(c) = K\Phi_f I_0\, e^{-\mu_{\text{em}}l_y}\left(e^{-\mu_{\text{ex}}x_{1j}} - e^{-\mu_{\text{ex}}x_{2j}}\right) \tag{13}$$

Equivalently, when concentration is expressed in mg/L and position in mm, the response can be written as

$$I_{f,j}(c) = F\left(e^{-\beta c x_{1j}} - e^{-\beta c x_{2j}}\right) \tag{14}$$

where $F$ contains the concentration-independent throughput and the constant lateral emission attenuation, and $\beta$ converts $\varepsilon_{\text{ex}}$ to the chosen concentration and length units. The difference of exponentials explains why both receiver endpoints matter [37]. At low concentration, Equation 14 reduces to $I_{f,j}$ approximately $F\beta c(x_{2j} - x_{1j})$, but at larger $c$ its derivative changes sign. Window position and width therefore set the response maximum, photon collection, and the concentration at which a channel enters its anomalous regime.

## EXPERIMENTS AND COMPUTATIONAL METHODS

### Fluorophore and fluorometry

L-Tryptophan was used as a model endogenous fluorophore in the experiment. Aqueous standards were prepared for spatial-profile fitting and for full-range inversion. Absorbance and fluorescence spectra established the excitation and emission bands shown as the inset in Figure 1A. The custom lateral 90-degree collection assembly comprised a fixed excitation axis, an extended 10 mm optical cell, imaging optics, precision translation stages, and aperture-defined receiver windows. Coordinates were referenced to the excitation entrance face as shown in Figure 1B. More details about the experimental setup and the reagents are organized in Supporting Information Section S2. Figure 1C shows fitted fluorescence line-density profiles from $10\ \text{mg/L}$ to $90\ \text{mg/L}$. Below approximately $10\ \text{mg/L}$ the emission field over the analytical region is comparatively uniform, whereas increasing concentration shifts useful fluorescence toward the entrance and steepens spatial attenuation of the excitation [7]. Figure 1D integrates these fields over 2 mm receiver windows. Moving a receiver toward the entrance delays the response maximum but does not preserve global linearity, each isolated channel eventually reaches vanishing-gradient boundary and then decreases into the anomalous regime [7]. The effect is therefore geometric redistribution, not simply a scalar loss that can always be recovered by multiplying by an absorptive correction factor. The geometry screen and simulations used a $0-125\ \text{mg/L}$ domain and the experimentally motivated noise model detailed in the Supporting Information Section S4. Experimental validation comprised 25 concentrations from $5\ \text{mg/L}$ to $125\ \text{mg/L}$ at $5\ \text{mg/L}$ increments, each measured in five replicates ($n = 125$ per configuration). Model parameters and raw receiver values were not linearized before inversion.

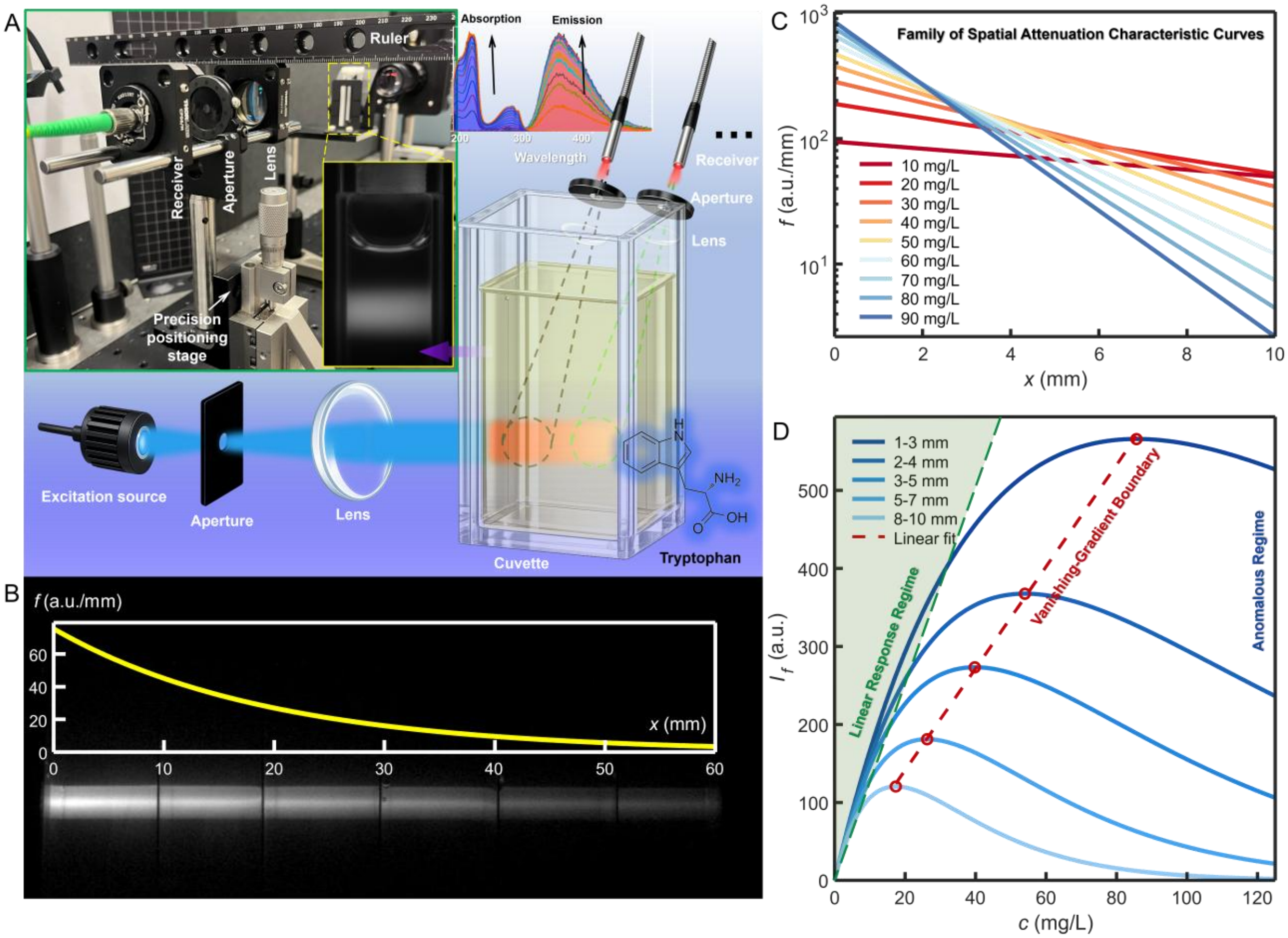

**Figure 1. Spatial origin and geometric encoding of the inner filter effect.** (A) Multireceiver $\mathbf{90}$-degree fluorometric arrangement with aperture-defined observation windows and tryptophan absorption/emission spectra. (B) Extended-cell image and fitted exponential line-density profile. (C) Fitted fluorescence line-density profiles for $\mathbf{10-90\ mg/L}$ tryptophan. (D) Integrated responses of $\mathbf{2\ mm}$ windows at different receiving positions, open red circles identify stationary points.

## Multichannel Inversion

In the lateral fluorescence-collection arm shown in Figure 1A, multiple spatially separated receivers were introduced. The fluorescence intensities recorded by these receivers were then used in a nonlinear optimization framework to iteratively recover the analyte concentration. As illustrated in Figure 2A, consider an arbitrary true tryptophan concentration $c_0$ within the full measurement range. The corresponding multireceiver fluorescence measurements form the joint fluorescent intensity vector

$$\mathbf{I}_{\text{meas}} = \begin{bmatrix} I_1(c_0) \\ I_2(c_0) \\ \vdots \\ I_j(c_0) \end{bmatrix} = \begin{bmatrix} I_{01} \\ I_{02} \\ \vdots \\ I_{0j} \end{bmatrix} \tag{15}$$

Given the concentration estimate $c_i$ at the $i$-th iteration, the forward model (Equation 14) predicts

$$\mathbf{I}(c_i) = \begin{bmatrix} I_1(c_i) \\ I_2(c_i) \\ \vdots \\ I_j(c_i) \end{bmatrix} = \begin{bmatrix} I_{i1} \\ I_{i2} \\ \vdots \\ I_{ij} \end{bmatrix} \tag{16}$$

The residual in the $j$-th receiver channel is

$$\delta_{ij} = I_{ij} - I_{0j} \tag{17}$$

The residuals from all receiver channels are combined into the scalar objective function

$$\Delta(c_i) = \frac{1}{2}\sum_{j=1}^{m}\left|\delta_{ij}\right|^2 = \frac{1}{2}\sum_{j=1}^{m}\left|\mathbf{I}(c_i) - \mathbf{I}_{\text{meas}}\right|^2 \tag{18}$$

The multichannel nonlinear inversion seeks the concentration estimate that minimizes this objective function:

$$\hat{c} = \arg\min_{c_i} \Delta(c_i) \tag{19}$$

such that, under ideal noise-free conditions, $\hat{c}$ converges to the true concentration $c_0$. Newton, gradient-descent, and Levenberg-Marquardt solvers were bounded to the physical interval, whose update equations and safeguards are provided in Supplementary Section S8.

Figure 2B examines how increasing the number of receiver channels affects the maximum-normalized objective-function landscape. The light-gray and dark-gray contours represent objective-function levels of $10^{-2}$ and $10^{-4}$, respectively, after normalization by the global maximum. A single receiver spanning $2-4$ mm produces an off-diagonal low-residual branch because the same intensity occurs on the increasing and decreasing sides of its calibration curve. This reflects the two-valued concentration inversion caused by the inner filter effect in fluorescence quantification. Moreover, near the vanishing-gradient point, the error contours extend far from the true solution. Consequently, even a small fluorescence-intensity measurement error can produce a large concentration-prediction error for the analyte, indicating both reduced sensitivity and poor robustness to measurement noise.

Introducing additional receiver channels with a displaced stationary point on the fluorescence-collection side substantially alleviates these quantitative limitations induced by the inner filter effect. A two-channel configuration is already sufficient to eliminate the two-valued ambiguity and make the concentration inversion globally unique. Nevertheless, near the vanishing-gradient boundary, the quantitative result remains relatively sensitive to fluorescence-intensity measurement errors. Further increasing the number of receiver channels improves both response sensitivity and robustness against measurement uncertainty. This distinction between global uniqueness and local slope is central: extra channels first remove competing solutions and only then improve precision. However, the improvement exhibits diminishing returns, and the quantitative performance in the nonlinear high-concentration regime remains inferior to that in the approximately linear low-concentration regime. Objective-function landscapes for additional receiver geometries are provided in Supplementary Note S3 in detail.

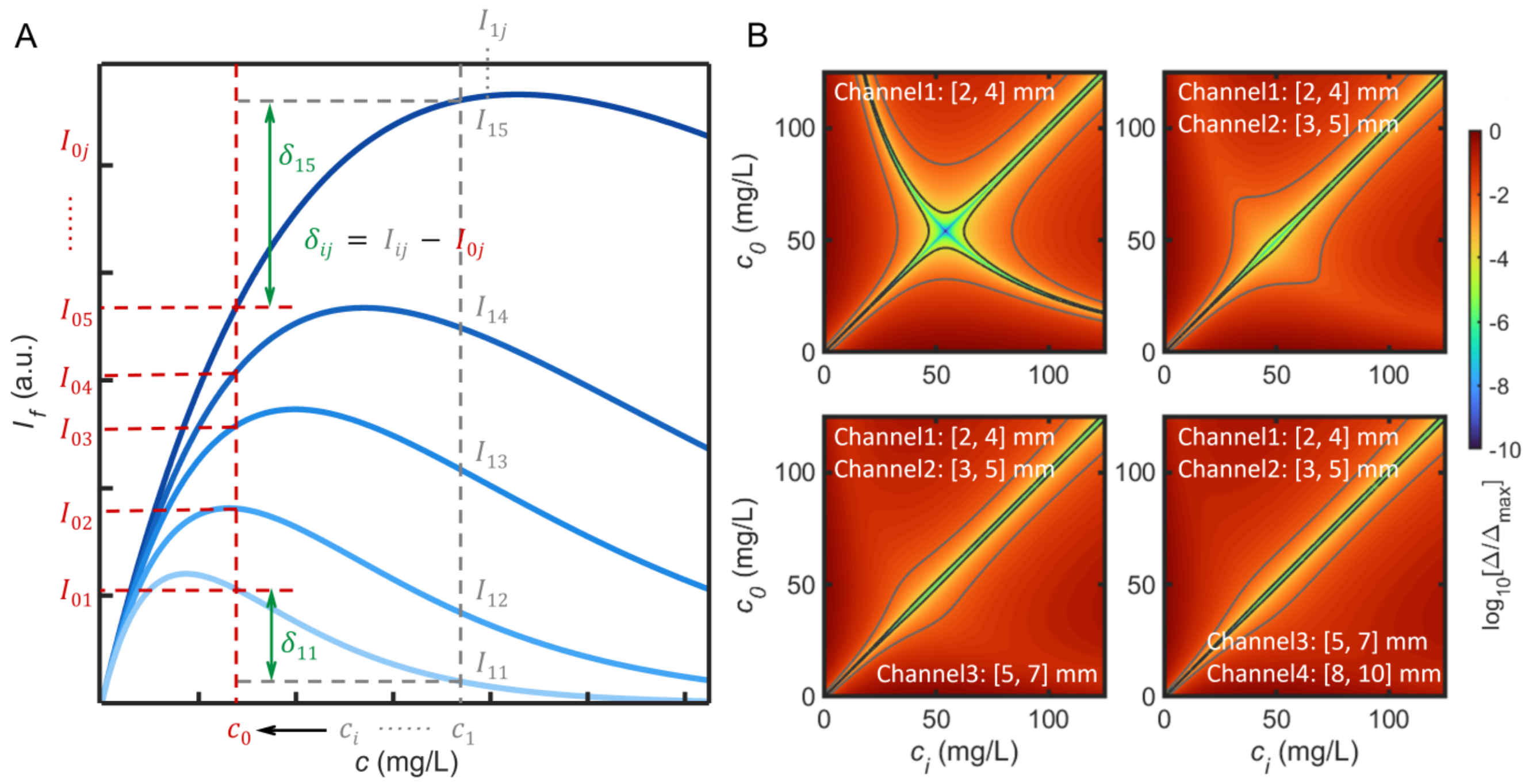

**Figure 2. Multireceiver nonlinear concentration inversion.** (A) Channel residuals between a true concentration $\boldsymbol{c_0}$ and trial concentration $\boldsymbol{c_i}$ during iteration. (B) Log-normalized objective surfaces for the fluorometric configurations with one to four receiver channels. The single-channel surface contains an off-diagonal ambiguity, spatially complementary channels collapse the low-residual set toward $\boldsymbol{c_i = c_0}$. Light- and dark-gray contours denote representative normalized error levels of $\mathbf{10^{-2}}$ and $\mathbf{10^{-4}}$.

## RESULTS AND DISCUSSION

### Uncertainty and Sensitivity Controlled by Receiver Geometry

Nonlinear inversion converts intensity noise into a concentration-dependent and potentially asymmetric uncertainty. We therefore used a profile-likelihood construction to evaluate anti-interference capability rather than a symmetric slope-only error bar. For each true concentration $c_0$, the connected set satisfying $\Delta\chi^2 \leq 3.8415$ defines the approximate 95% interval for one fitted parameter. The threshold always has one degree of freedom even when several receiver channels are measured, because only the scalar concentration is estimated. Profile methods are particularly useful near weak curvature or asymmetric likelihoods and provide a direct diagnostic of practical nonidentifiability [38, 39]. The 95% error half-width $E_{95}(c_0)$, endpoint treatment, and disconnected-set rule are detailed in Supplementary Section S4. The full-range-averaged uncertainty used for geometry screening is

$$\overline{E_{95}} = \frac{1}{c_{\max}-c_{\min}} \int_{c_{\min}}^{c_{\max}} E_{95}(c_0)\, dc_0 \tag{20}$$

A complementary local metric is the noise-normalized sensitivity $S_N(c_0)$, equivalent to the square root of the scalar Fisher information under independent Gaussian noise. Unlike a signed average derivative, it cannot be artificially canceled when one channel is increasing and another decreasing. Its range average is

$$\overline{S_N} = \frac{1}{c_{\max}-c_{\min}} \int_{c_{\min}}^{c_{\max}} S_N(c_0)\, dc_0 \tag{21}$$

Locally, $E_{95}$ is approximately $1.96/S_N$ (Equation S24), but the equality fails when the objective is asymmetric or clipped by a domain boundary. Accordingly, the two metrics answer related but nonidentical design questions: noise-normalized sensitivity describes local derivative-to-noise ratio, whereas the 95% error half-width reflects the uncertainty of the inverse problem.

Thirty-five feasible two-receiver geometries were evaluated with the same spatial forward model and noise law (Supplementary Section S5). Mean $E_{95}$ ranged from $0.884$ to $1.982\ \mathrm{mg/L}$ and mean $S_N$ from $1.652$ to $3.832\ \mathrm{a.u./(mg/L)}$, nearly a $2.2$-fold span in uncertainty. Geometry is therefore a first-order analytical design variable. Configuration 15, $[0, 1]$ and $[4, 7]$ mm, minimized mean $E_{95}$ at $0.884\ \mathrm{mg/L}$. Its narrow entrance window samples the steep concentration-dependent redistribution where photons remain available at high optical density, while the broader downstream window retains a response shape that is sufficiently different to prevent a second solution. Configuration 34, $[5, 6]$ and $[8, 9]$ mm, performed worst by both metrics (mean $E_{95} = 1.982\ \mathrm{mg/L}$; mean $S_N = 1.652\ \mathrm{a.u./(mg/L)}$). Both windows are deep in the attenuated region, so their signals are weak and their derivatives become similar, adding such channels increases data volume but little independent information. Configuration 21, $[0, 3]$ and $[4, 7]$ mm, maximized mean $S_N$ at $3.832\ \mathrm{a.u./(mg/L)}$ and retained mean $E_{95} = 0.978\ \mathrm{mg/L}$. Broadening the entrance receiver collects more photons and improves local derivative-to-noise ratio, but it also spatially averages the most rapidly varying part of the field. This explains why the maximum-$S_N$ and minimum-$E_{95}$ designs are close but not identical. Across all 35 designs, mean $E_{95}$ and mean $S_N$ were strongly anticorrelated (Pearson $r = -0.761$), yet the remaining scatter is scientifically informative because it marks nonquadratic or asymmetric objective structure.

A constrained continuous genetic algorithm independently reproduced the screen as detailed in Supplementary Section S6. Minimization of mean $E_{95}$ gave $[0, 0.7]$ and $[4.2, 7.2]$ mm with $0.876\ \mathrm{mg/L}$; maximization of mean $S_N$ gave $[0, 3.0]$ and $[4.2, 7.2]$ mm with $3.836\ \mathrm{a.u./(mg/L)}$. The downstream window converged to nearly the same interval in both searches, whereas the entrance-window width reflected the selected metric. This agreement indicates a broad and physically interpretable optimum rather than an isolated numerical artifact.

Figures 3C and 3D resolve the concentration dependence of the selected metric. A comprehensive analysis of the concentration-resolved uncertainty and sensitivity for all investigated receiver geometries is presented in Supplementary Section S7. Precision is strongest in the quasi-linear low-concentration region and degrades where response vectors flatten. At high concentration, pIFE compresses useful excitation into an increasingly shallow entrance layer, therefore downstream channels then lose photons and derivative information. Geometry can redistribute and combine the remaining information but cannot create photons, explaining both the advantage of an entrance window and the diminishing benefit of additional receivers. The result also clarifies the relationship to front-face fluorescence: front-face collection reduces the effective path, whereas the present method deliberately pairs an entrance-region measurement with a distinct downstream integral so that spatial contrast, not merely signal magnitude, identifies concentration [10, 34].

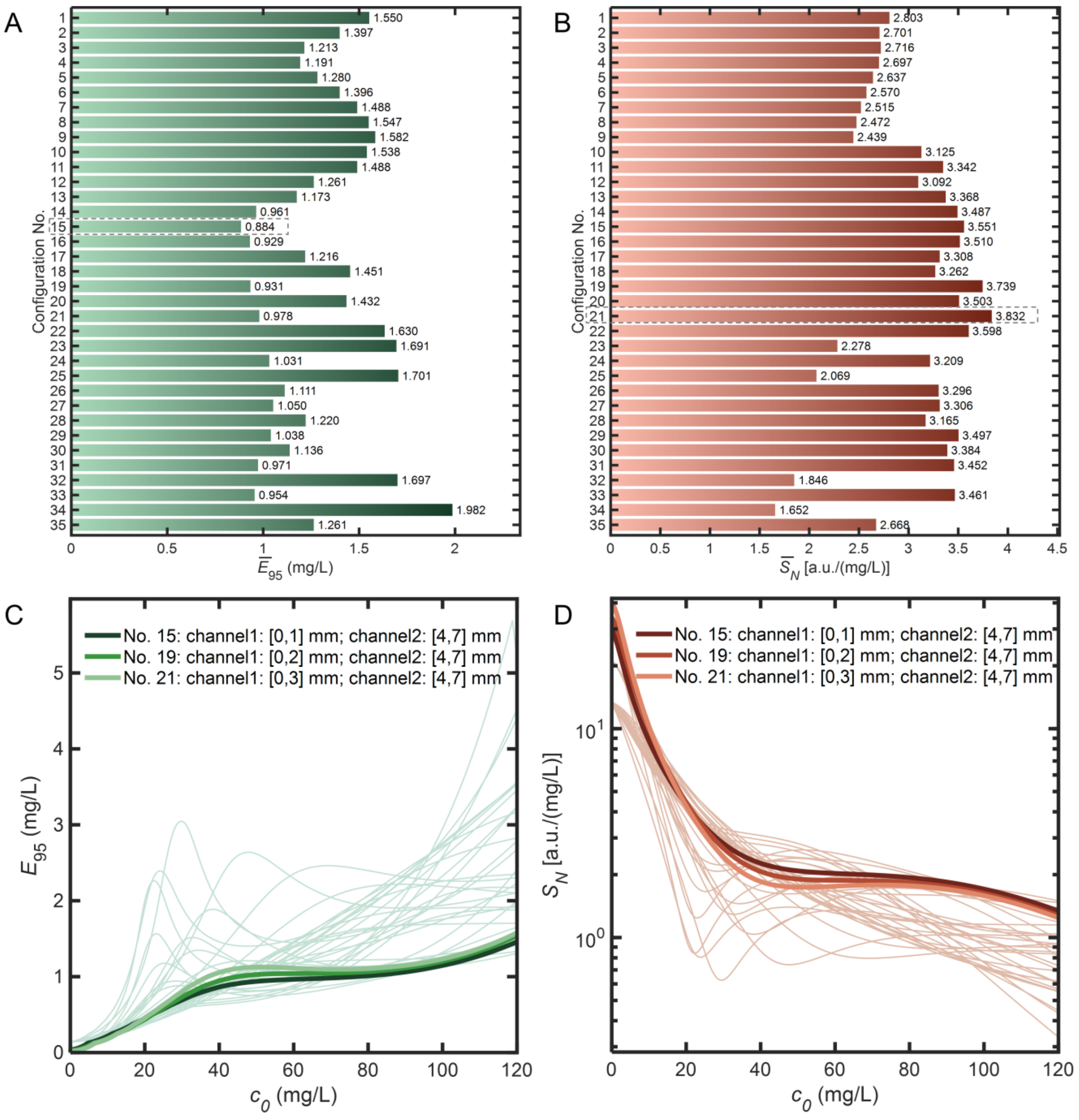

**Figure 3. Geometry optimization for two-receiver fluorometric configuration.** (A) Mean profile $95\%$ concentration-error half-width and (B) mean noise-normalized sensitivity for $35$ configurations. Dashed boxes mark the best discrete geometric configurations. (C,D) Concentration-resolved $\boldsymbol{E_{95}(c_0)}$ and $\boldsymbol{S_N(c_0)}$, highlighted curves correspond to near-optimal designs in average.

## Nonlinear Optimization Solvers

In this work, as mentioned above, Newton, gradient-descent, and Levenberg–Marquardt iterations were evaluated as representative nonlinear solvers. Descriptions of these algorithms and their corresponding update equations and safeguards are provided in Supplementary Information S8. Each solver was initialized from both the low-concentration and high-concentration ends of the investigated range. Figure 4 tracks the iterative trajectories of the channel predictions, the reduction in residuals, and the absolute concentration error. Once the complementary receiver channels eliminated the competing minimum, all three methods converged to the same terminal concentration estimate, irrespective of the initialization point. The nonlinear inversion is not restricted to these three representative algorithms. In principle, any suitable nonlinear solver that successfully converges to the unique global minimum should recover the same accurate concentration estimate, although the convergence rate, numerical stability, and computational cost differ among algorithms.

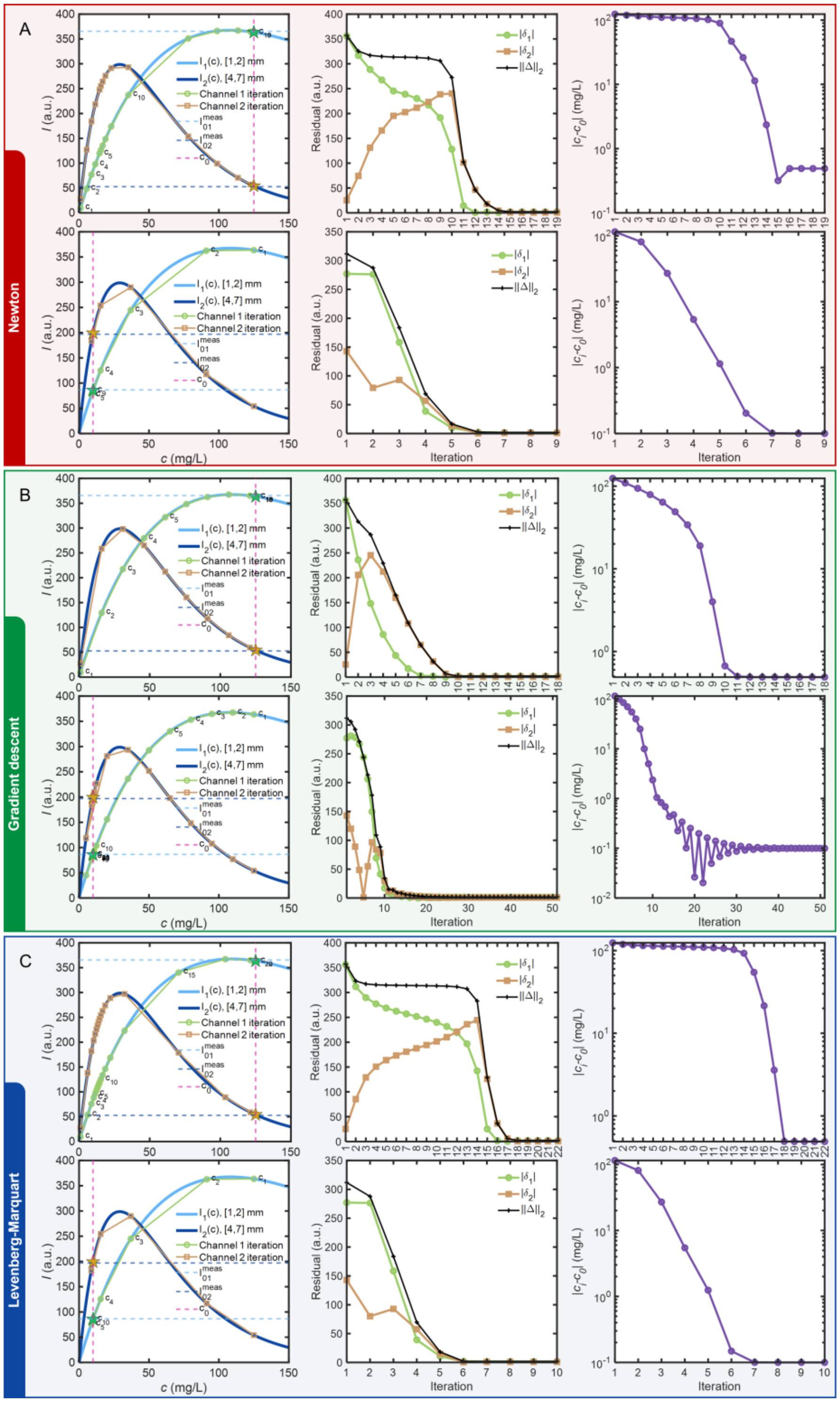

**Figure 4. Consistent convergence of Newton, gradient-descent, and Levenberg-Marquardt inversion for low-to-high and high-to-low initializations.** Left: the first column shows the evolution of the concentration estimate $c_i$ over successive iterations, progressing from the initial estimate $c_1$ toward the final solution marked by a star. Center: the progressive reduction of the channel residuals and the scalar

objective function during iteration. Right: absolute analyte concentration error. All methods converge to the same solution with complementary receivers.

The algorithms nevertheless exhibit distinct numerical behavior. Newton iteration is rapidly convergent near a well-conditioned minimum but can take an excessive step when curvature is small or changes sign (Figure 4A). Bounded steps and damping are therefore necessary near channel stationary points. Gradient descent requires only first-order information and is robust to imperfect curvature, but it progresses slowly along a shallow valley and can oscillate if the step size is too large (Figure 4B). Levenberg-Marquardt exploits the Jacobian structure while damping poorly conditioned Gauss-Newton steps, giving the most balanced convergence for this least-squares problem (Figure 4C) [40]. The decisive observation is that no local solver can repair a genuinely noninjective single-channel calibration. Only multireceiver geometry can remove the ambiguity and the optimizer then determines computational speed and robustness.

## Full-Range Quantification

The full-range nonlinear quantification strategy was tested with $125$ measurements spanning $5 - 125\,\mathrm{mg/L}$. A complete presentation of the experimental results is provided in Supplementary Section S9. For receivers $[1, 2]$ and $[4, 7]$ mm, the three nonlinear solvers returned identical estimates to the reported precision. Regression against reference concentration was $y = 1.004x - 0.291$ with $R^2 = 0.9959$. Agreement of the slope with unity and the near-zero global bias show that the spatial forward model crosses the stationary point and recovers the high-concentration decreasing regime rather than fitting a local monotonic segment. A third receiver at $[3, 6]$ mm further improved the regression to $y = 1.004x - 0.358$ with $R^2 = 0.9972$. RMSE decreased by $17.1\%$ to $1.934\,\mathrm{mg/L}$, MAE decreased by $20.2\%$ to $1.440\,\mathrm{mg/L}$, and the $95$th percentile absolute error decreased from $4.920\,\mathrm{mg/L}$ to $3.706\,\mathrm{mg/L}$. The improvement is consistent with increased Fisher information and redundancy. It is smaller than the jump from one to two channels because the second channel establishes global uniqueness, whereas the third mainly sharpens an already unique minimum.

Remarkably, errors remain concentration dependent. Larger excursions occur where one or more channel derivatives are small and where intensity-dependent noise is larger. This heteroscedasticity is expected from Equation 17 and is obscured by $R^2$ alone. The maximum absolute errors were $7.854$ and $6.446\,\mathrm{mg/L}$ for two and three receivers, respectively, so reporting RMSE, MAE, tail error, and concentration-resolved residuals is more informative than a single regression coefficient. The present experiment used aqueous tryptophan and therefore establishes the optical inverse-problem principle, not yet a clinical assay. Blood, serum, and protein formulations add scattering, binding-dependent quantum yield, and spectral shifts. In such matrices, IFE must be distinguished from true self-quenching, aggregation, energy transfer, and matrix fluorescence [19, 31]. These effects could be accommodated by extending the forward model or by introducing additional spectral channels. Nevertheless, systematic calibration of the inner filter effect remains the essential foundation for any more comprehensive study before clinical translation.

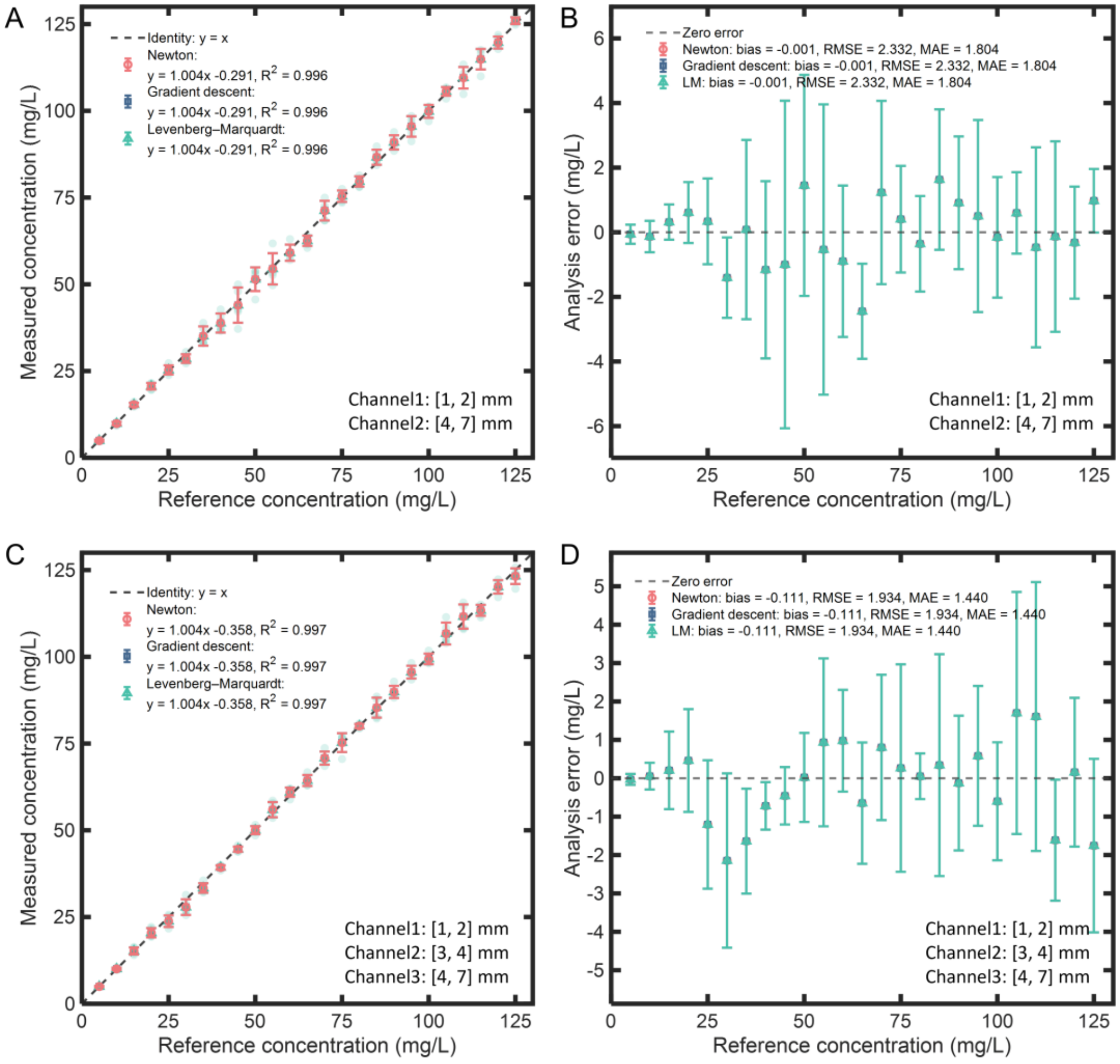

**Figure 5. Experimental full-range quantification of tryptophan fluorescence.** (A, B) Regression and concentration-resolved analysis errors for two receivers. (C, D) Corresponding results for three receivers. Individual measurements for each sample and concentration-level summaries are shown, Newton, gradient-descent, and Levenberg-Marquardt estimates overlap.

## CONCLUSIONS

The IFE makes a single fluorescence intensity an unreliable concentration coordinate once the response becomes nonmonotonic. This work replaces that scalar coordinate with a geometry-encoded multichannel trajectory. Spatially separated receiver windows integrate different portions of the attenuation-shaped emission field [37], restoring global uniqueness. Profile uncertainty and noise-normalized sensitivity then provide complementary criteria for determining the optimal geometry of the receiver endpoints. Two receivers removed the two-valued inversion over $5 - 125\ \mathrm{mg/L}$, and a third reduced RMSE and MAE while showing the expected diminishing return. The strategy differs fundamentally from post-hoc IFE correction: it retains spatial attenuation as information and estimates concentration directly from the nonlinear spatial forward model. The immediate next steps are rigorous matrix validation, joint treatment of scattering and sIFE, propagation of calibration-parameter uncertainty, and hardware implementation with fixed multielement detectors [34]. More generally, the results demonstrate that observation geometry can be designed and analyzed as part of the metrological model whenever attenuation redistributes optical emission in space.

## ASSOCIATED CONTENT

### Supporting Information

Supporting Information contains complete collection-geometry derivations; instrumentation and reagent preparation; objective-function maps; profile-likelihood and sensitivity definitions; all 35 screened geometries; genetic-algorithm introductions and settings; concentration-resolved uncertainty and sensitivity; nonlinear-solver iteration equations and relevant introductions; complete two-receiver and three-receiver measurement data.

## ACKNOWLEDGMENTS

This work was supported by the National Natural Science Foundation of China (No. 62405177). We also acknowledge support from the Outstanding Doctoral Graduates Development Scholarship of Shanghai Jiao Tong University.